\documentclass[aps,prb,amssymb,twocolumn, superscriptaddress
]{revtex4-2}

\usepackage{graphicx}
\usepackage{amsmath}
\usepackage{xcolor}

\usepackage{glossaries-extra}
\setabbreviationstyle[acronym]{long-short}
\makeglossaries
\newacronym{sot}{SOT}{spin--orbit torque}
\newacronym{saf}{SAF}{synthetic antiferromagnet}
\newacronym{pma}{PMA}{perpendicular magnetic anisotropy}
\newacronym{stt}{STT}{spin transfer torque}
\newacronym{vcma}{VCMA}{voltage control of magnetic anisotropy}

\usepackage[pagewise]{lineno}
\usepackage[version=4]{mhchem}

\newcommand{\mh}[1]{\textcolor{black}{#1}}

\begin{document}

\title{Pulse Shaping Strategies for Efficient  Switching of Magnetic Tunnel Junctions by Spin-Orbit Torque}

\author{Marco Hoffmann}
\email[]{Marco.Hoffmann@mat.ethz.ch}
\affiliation{Department of Materials, ETH Zurich, 8093 Zurich, Switzerland}

\author{Viola Krizakova}

\affiliation{Department of Materials, ETH Zurich, 8093 Zurich, Switzerland}

\author{Vaishnavi Kateel}
\affiliation{IMEC, Kapledreef 75, 3001 Leuven, Belgium}
\affiliation{Department of Electrical engineering ESAT, KU Leuven, Kasteelpark Arenberg 10, 3001 Leuven, Belgium}

\author{Kaiming Cai}
\affiliation{IMEC, Kapledreef 75, 3001 Leuven, Belgium}

\author{Sebastien Couet}
\affiliation{IMEC, Kapledreef 75, 3001 Leuven, Belgium}

\author{Pietro Gambardella}
\email[]{pietro.gambardella@mat.ethz.ch}
\affiliation{Department of Materials, ETH Zurich, 8093 Zurich, Switzerland}

\date{\today}

\begin{abstract}

The writing energy for reversing the magnetization of the free layer in a magnetic tunnel junction (MTJ) is a key figure of merit for comparing the performances of magnetic random access memories with competing technologies. Magnetization switching of MTJs induced by spin torques typically relies on square voltage pulses. Here, we focus on the switching of perpendicular MTJs driven by spin--orbit torque (SOT), for which the magnetization reversal process consists of sequential domain nucleation and domain wall propagation. By performing a systematic study of the switching efficiency and speed as a function of pulse shape, we show that shaped pulses achieve up to 50\% reduction of writing energy compared to square pulses without compromising the switching probability and speed. Time-resolved measurements of the tunneling magnetoresistance reveal how the switching times are strongly impacted by the pulse shape and temperature rise during the pulse. The optimal pulse shape consists of a preheating phase, a maximum amplitude to induce domain nucleation, and a lower amplitude phase to complete the reversal. Our experimental results, corroborated by micromagnetic simulations, provide diverse options to reduce the energy footprint of SOT devices in magnetic memory applications. 
\end{abstract}

\maketitle

\section{Introduction}
Current-induced \gls{sot} switching of magnetic tunnel junctions (MTJs) has been investigated to develop fast and low-energy magnetic random access memories (MRAM) with high endurance and speed \cite{Manchon2019, Liu2012, Pai2012, Cubukcu2014, Aradhya2016, Fukami2016, Cubukcu2018, Garello2018, Grimaldi2020, Endoh2020a, Krizakova2021}. In MTJ devices, the magnetization of the free layer can be manipulated by electrical currents passing either through the junction or an adjacent heavy metal layer, inducing \gls{stt} \cite{Ralph2008, Brataas2012} or \gls{sot} \cite{Miron2011, Garello2013}, respectively. SOT-induced magnetization reversal reliably reaches the sub-nanosecond regime \cite{Garello2014, Grimaldi2020, Krizakova2020, Wu2021a}.
However, the energy required for writing information in MTJs is a drawback for emerging SOT technologies \cite{Krizakova2022, Vedmedenko2020}. Therefore, dual pulsing schemes have been explored to decrease the switching energy consumption in MTJs, e.g., by the simultaneous application of SOT, \gls{stt}, and \gls{vcma} \cite{Grimaldi2020, Pathak2020, Wu2021a, Krizakova2021}. 

The impact of self-heating due to Joule dissipation on the switching process has also been investigated \cite{Grimaldi2020, Krizakova2021, Chavent2016, Mihajlovic2020, Goto2022, Hadamek2023}. The temperature increase $\Delta T$ of the magnetic free layer significantly reduces its \gls{pma} and thus the critical current density \cite{Yamaguchi2005, You2006, Papusoi2008, Hadamek2022}. 
In pulsed switching, $\Delta T$ is assumed to exponentially approach its saturation value, which is proportional to the applied electrical power density \cite{Sousa2004, Kim2008, Fukami2013}. The resultant decrease in \gls{pma} reduces the nucleation time of a domain with inverted magnetization in the free layer of the MTJ \cite{Grimaldi2020, Strelkov2018}, which is the critical step to trigger SOT switching  \cite{Baumgartner2017, Sala2023, Martinez2015}. 
This insight suggests that the writing pulse may be shaped so as to optimally drive the different phases of the magnetization reversal process.

Previous works suggest that non-conventional pulsing schemes can be advantageous for several reasons. For instance, triangular pulse shapes can reduce the risk of degradation in resistive random access memory, thanks to the improved control of current overshoots \cite{Lee2015}. 
Micromagnetic simulations of STT switching of MTJs show that a current spike in the beginning of the pulse provides larger spin torque and Oersted field, such that both the switching speed and energy can be improved compared to conventional square pulses \cite{Pathak2017}. Similarly, micromagnetic simulations of SOT switching predict that using a high-amplitude pulse to initiate domain nucleation followed by a longer low-amplitude pulse for domain wall propagation can reduce the switching energy \cite{Jin2022}. Switching experiments of perpendicular MTJs have shown that a reduction of the \gls{pma} during the first half of the SOT pulse leads to a concomitant reduction of the critical current \cite{Wu2021a}. Moreover, micromagnetic simulations of SOT-assisted STT switching in three-terminal MTJs yield faster and more energy-efficient magnetization reversals, if an SOT pulse is applied before the main STT pulse, such that it tilts the magnetization away from the easy axis direction \cite{Pathak2020, Cai2021, Yoshida2022}. \mh{Adapting the rise and fall time of pulses can also be advantageous to improve the reliability of precessional switching induced by VCMA switching \cite{Yamamoto2019}.}
However, most experimental studies of MTJ switching so far focused on the application of square voltage pulses with a constant amplitude throughout the pulse length.

Here, we report on a systematic investigation of SOT switching of three-terminal MTJs employing diverse pulse shapes. Post-pulse and time-resolved measurements reveal the switching dynamics and the critical ranges for switching voltage, energy, and pulse duration. We find that by allowing larger voltages in certain segments of the pulse, the total switching energy can be reduced by up to about 50\% with respect to square pulses \mh{without significantly affecting the switching speed. Measurements performed using square, triangular, sine, and spiked voltage pulses reveal that the energy efficiency of magnetization reversal can be minimized by shaping the pulses such that the SOT and $\Delta T$ peak at approximately the same time to initiate domain nucleation while leaving enough time for the pulse to complete the reversal by SOT-driven domain wall motion. Micromagnetic simulations that take into account the time-dependent self-heating of the free layer corroborate the experimental results.}

This paper is organized as follows. The samples, experimental setup and simulation methods are described in Section \ref{2}. Section \ref{3} elucidates the energy efficiency of different voltage pulse shapes employed for SOT switching. Subsequently, the importance of different time segments of the voltage pulse is addressed in Section \ref{4}. The results from our micromagnetic simulations are presented in Section \ref{5}.

\section{\label{2}Samples and methods}

\subsection{Devices} 
\begin{figure}
\includegraphics[width=85mm]{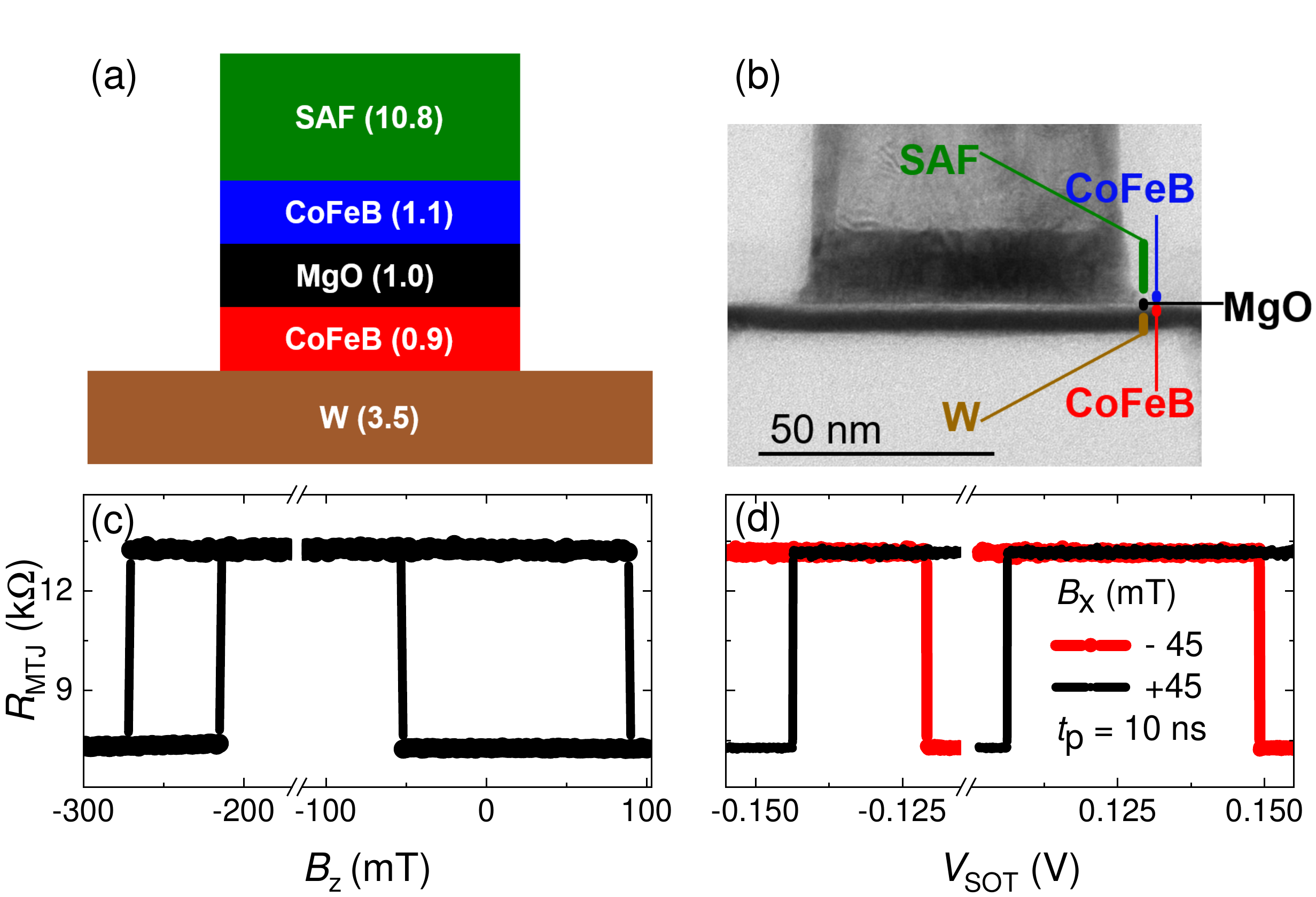}
\caption{\label{fig0:sample_charac} \mh{(a) Schematic of the MTJ stack including, from bottom to top, the $\beta$-W SOT track, the \ce{Co20Fe60B20} magnetic free layer, the MgO insulating layer, the \ce{Co_{17.5}Fe_{52.5}B30} magnetic reference layer, a W spacer layer and the synthetic antiferromagnet made of Pt/Co layers. Numbers indicate thickness in nanometer. (b) Exemplary transmission electron microscopy image of a 60\,nm wide MTJ device. (c) TMR major loop of a representative MTJ device as a function of $B_\text{z}$. (d) TMR minor loops as a function of SOT voltage for both signs of $B_\text{x}$ measured on the same device  using 10-ns-long voltage pulses. }
}
\end{figure}

\mh{The MTJs investigated in this work have a circular cross section with a diameter of 80\,nm and comprise a free layer of \ce{Co20Fe60B20} (0.9\,nm) at the bottom, an MgO tunnel barrier (1.0\,nm), and a reference layer of \ce{Co_{17.5}Fe_{52.5}B30} (1.1\,nm) pinned from the top by a synthetic antiferromagnetic stack made of Pt/Co multilayers. 
The SOT track underneath the free layer is made of $\beta$-W with resistance $R_\text{SOT} \approx 450\, \Omega$. The structure and cross section of the MTJ stack are presented in Fig.~\ref{fig0:sample_charac}(a) and (b), respectively. More details on device fabrication are given in Refs.~\onlinecite{Garello2018, Garello2019}. Figure~\ref{fig0:sample_charac}(c) and (d) show the tunneling magnetoresistance (TMR) of a representative device as a function of out-of-plane magnetic field $B_{z}$ and 10-ns-long SOT voltage pulses of amplitude $V_\text{SOT}$. The resistance of the representative junction, $R_\text{MTJ}$, varies between 7.2 and 13.0~k$\Omega$ in the parallel (P) and antiparallel (AP) state, respectively.} 

\begin{figure}
\includegraphics[width=85mm]{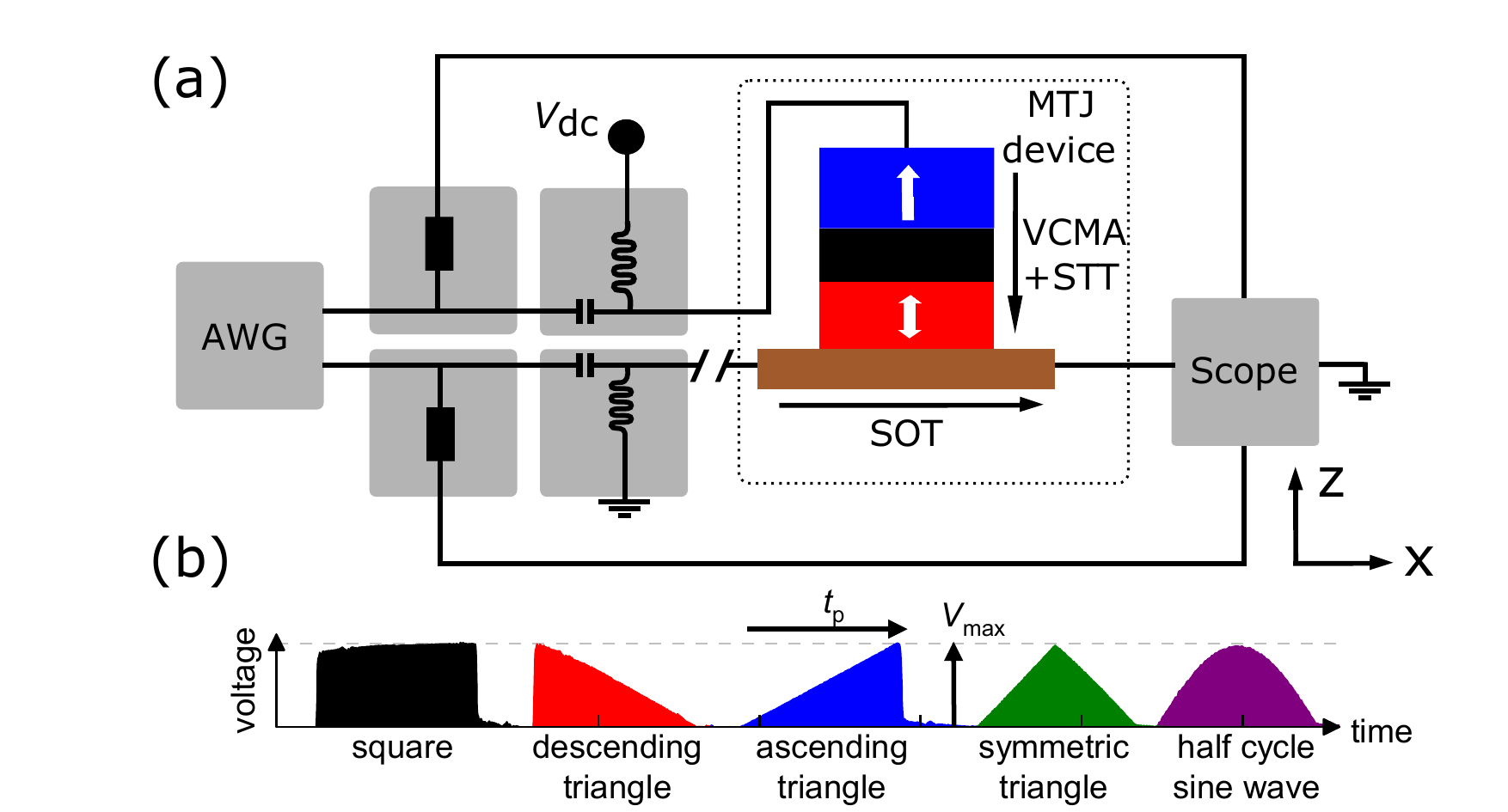}
\caption{\label{fig1:Exp} (a) Schematic of the experimental setup: An arbitrary waveform generator outputs two independent voltage pulses that pass pick-off tees for raw signal investigation and triggering and bias tees before entering the input terminals of the MTJ simultaneously. The (raw) pick-off tee signals and the signal passing the MTJ are acquired on a real-time oscilloscope.
(b) Different pulse shapes employed in this study: Square (black), descending triangle (red), ascending triangle (blue), symmetric triangle (green), and a half cycle sine wave (purple). $t_\text{p}$ and $V_\text{max}$ indicate the total length and maximum amplitude of the voltage pulse, respectively.}
\end{figure}

\subsection{Electrical measurements}
\mh{In the following, we focus on the switching between the P and AP state of the MTJ, with a constant magnetic field  $B_{x} = -45\,$mT applied parallel to the current along the $x$-direction. This field provides the symmetry breaking required for deterministic \gls{sot} switching of magnetic layers with PMA \cite{Miron2011}. Figure~\ref{fig1:Exp}(a) provides an overview of the experimental setup, in which an arbitrary waveform generator (AWG) with a bandwidth of 5\,GHz is used to inject two separate voltage pulses into 50\,$\Omega$ matched transmission lines connected to the two input terminals of the MTJs. 
The output waveform is measured by an oscilloscope with 10 GHz bandwidth. We perform two types of experiments: post-pulse switching measurements, in which $R_\text{MTJ}$ is measured several milliseconds after the voltage pulse by applying a dc voltage of 20~mV -well below any critical switching voltage- to determine the success of the switching attempt, and time-resolved switching measurements, in which $R_\text{MTJ}$ is measured in real time during the application of the pulse~\cite{Grimaldi2020}. The typical relaxation time of the temperature in our MTJ devices is of the order of a few ns~\cite{Grimaldi2020, Krizakova2021}. Therefore, in either type of measurement, the temperature and magnetization are fully relaxed before every switching attempt.}

By applying simultaneously different voltages to the top terminal of the MTJ ($V_\text{STT}$) and the SOT track ($V_\text{SOT}$), the \gls{stt} and \gls{sot} biases can be controlled \cite{Grimaldi2020}. Specifically, if $V_\text{STT} \approx 0.5\,V_\text{SOT}$, the effective voltage drop across the MTJ ($V_\text{MTJ}$) vanishes and the switching is of pure \gls{sot} nature \cite{Grimaldi2020}. If $V_\text{MTJ} \ne 0$, the switching is affected by \gls{stt} and VCMA contributions. For time-resolved SOT switching, we apply a finite $V_\text{MTJ}$ in order to measure $R_\text{MTJ}$ during the pulse. This bias is about 30\% of the critical STT switching voltage and does not cause switching by itself, but alters the critical SOT switching conditions. Thus, we can detect the change of $R_\text{MTJ}$ and reveal the dynamics of the magnetization reversal in real-time, while maintaining SOT-dominated switching \cite{Grimaldi2020, Krizakova2021}.

The pulse shapes are affected by the limited bandwidth and stray capacitances of the electrical setup, which becomes most evident for pulse durations in the sub-nanosecond range. The 10\% to 90\% rise time of the AWG is 90\,ps. 
To make a fair comparison between pulse shapes with sharp features [Fig.~\ref{fig1:Exp}(b)] that are impacted differently by the finite bandwidth, we proceed in the following manner:  
Raw pulse shapes injected to the MTJ terminals are recorded on the oscilloscope using pick-off tees. Using these raw traces, the pulse energy $E_\text{p}$ is numerically calculated using Eq.~\eqref{equ_E-V} and the maximum voltage $V_\text{max}$ is read off.
$V_\text{max}$ is defined as the maximum applied voltage for each pulse as observed in the $50\,\Omega$ matched oscilloscope, whereas a larger voltage appears on the input terminals of the MTJ device due to the impedance mismatch \cite{Grimaldi2020}.

\subsection{Micromagnetic simulations}
We used the micromagnetic simulation software Mumax$^3$ \cite{Vansteenkiste2014} to model the SOT-induced switching by different pulse shapes. For this, we defined a magnetic cylinder of 80\,nm diameter and 1\,nm thickness in a simulation environment of ($84 \times 84 \times 1$)\, $\text{nm}^3$ discretized into 64 $\times$ 64 cells in the plane. 
The employed magnetic parameters at room temperature are the saturation magnetization $M_\text{sat}=1.1$\,MA/m, the \gls{pma} energy density $K_\text{u}=845\,\text{kJ/m}^3$, and the exchange stiffness $A_\text{ex}$=15\,pJ/m \cite{Grimaldi2020}. These parameters are varied with temperature for different pulse shapes, as explained below. The Curie temperature and the damping parameter 
are $T_\text{C}=750\,$K and $\alpha=0.1$, respectively.
The Dzyaloshinskii-Moriya interaction is parametrized by $D = 0.15\,$mJ/m$^2$, the in-plane magnetic field is $B_{x} = -45\,$mT, and we simulate the P-AP transition by starting with the magnetization initially pointing along $+z$ \cite{Grimaldi2020}.

\section{\label{3}Switching efficiency of different pulse shapes}
\label{section_switching_efficiency_of_pulse_shapes}
\begin{figure}
\includegraphics[width=85mm]{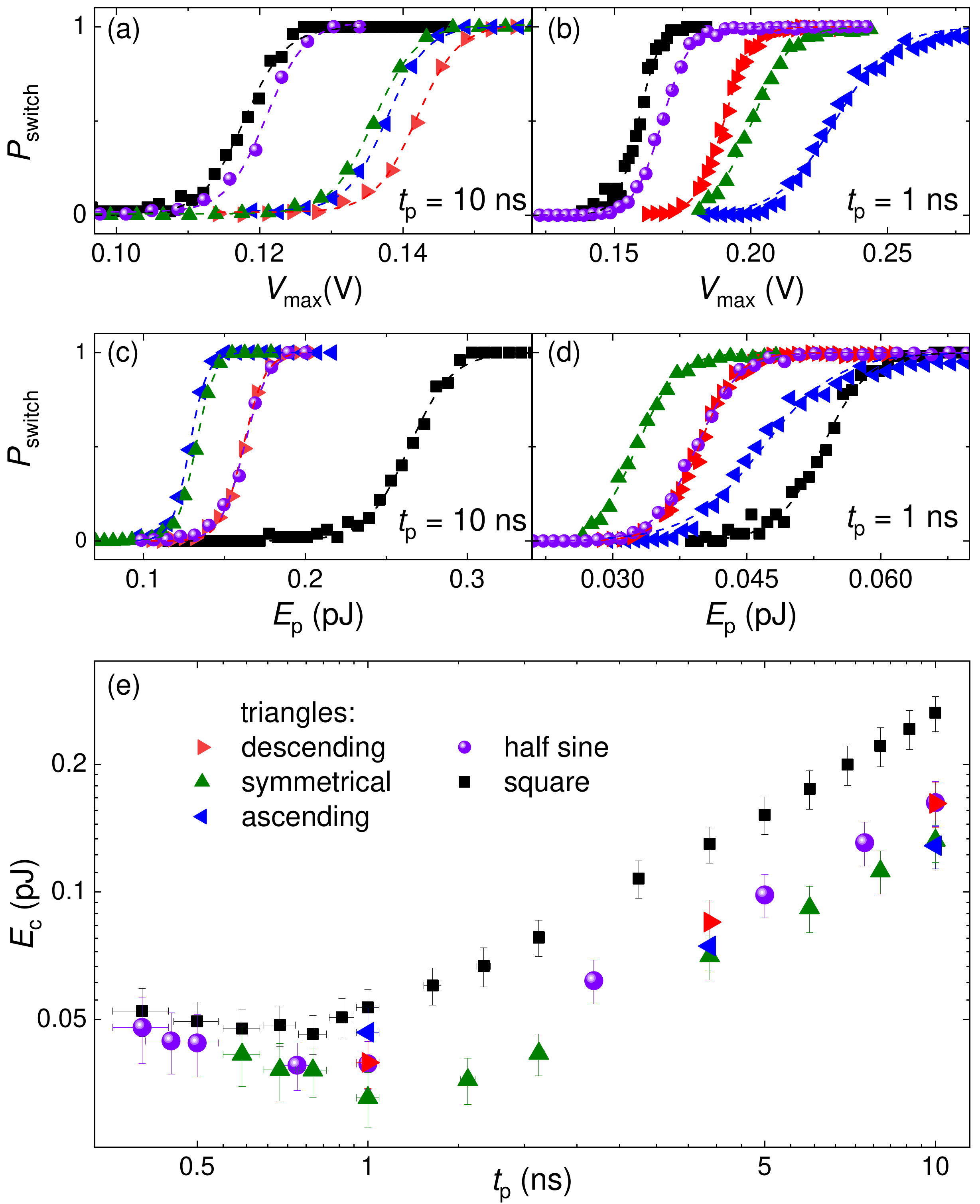}
\caption{\label{fig2:PP_shapes} 
SOT switching probability $P_\text{switch}$ for different pulse shapes at $t_\text{p} =$ 10~ns (a) and 1~ns (b) as a function of the maximum applied SOT voltage $V_\text{max}$ and the corresponding pulse energy $E_\text{p}$ in (c) and (d). 
(e) Critical switching energy $E_\text{c}$ as a function of $t_\text{p}$. All data were measured for the P-AP transition at $V_\text{MTJ}=0$.} 
\end{figure}

\subsection{Post-pulse measurements} We first study the SOT switching probability $P_\text{switch}$ of different pulse shapes as a function of $V_\text{max}$, keeping $t_\text{p}$ fixed to 10~ns or 1~ns and $V_\text{MTJ}=0$. These two pulse lengths are representative of different switching regimes, dominated by thermal activation and intrinsic angular momentum transfer, respectively \cite{Bedau2010, Garello2014}. The resulting sigmoidal curves are presented in Fig.~\ref{fig2:PP_shapes}(a) and (b). Clearly, keeping the same voltage for the entire pulse duration yields the most ``voltage-efficient'' switching, i.e.,~ switching is attained at the lowest $V_\text{max}$ for $P_\text{switch}=50\%$. However, plotting the same switching probabilities as a function of energy
\begin{equation}
\label{equ_E-V}
E_\text{pulse} = \int_{0}^{t_\text{p}} V_\text{pulse}^2(t) / R_\text{SOT} \, dt
\end{equation}
reveals a new perspective, shown in Fig.~\ref{fig2:PP_shapes}(c) and (d).
Here, the function $V_\text{pulse}(t)$ represents the pulse shape measured by the oscilloscope. 
For instance, for the descending triangular pulse [Fig.~\ref{fig1:Exp}(b)] $V_\text{pulse}(t)$ is close to the ideal shape
\begin{equation}
V_\text{pulse}(t) = V_\text{max} \cdot \left( 1 - t / t_\text{p} \right)
\end{equation}
for $t \in [0, t_\text{p}]$ and zero otherwise. 

\begin{figure}
\includegraphics[width=85mm]{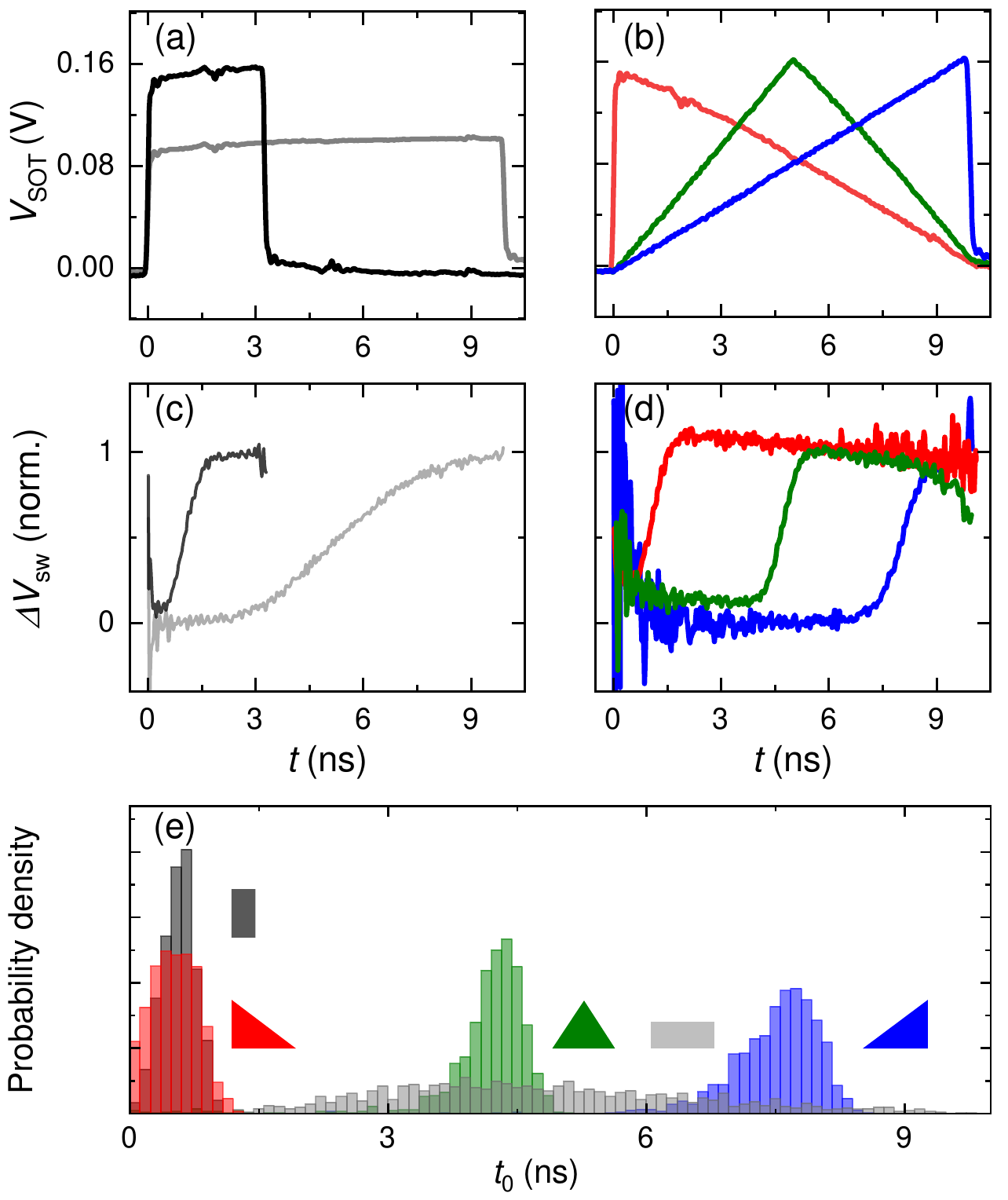} 
\caption{\label{fig3:TR_shapes} Pulse shapes supplied to the SOT terminal for (a) square and (b) triangular pulses. Time-resolved normalized switching traces of (c) square and (d) triangular pulses corresponding to the P-AP transition at $B_{x} = -45 $\,mT at $V_\text{MTJ} = 1.75\,V_\text{SOT}$ averaged over 500 individual switching traces. The triangular pulses with $t_\text{p}=10\,$ns as well as the shortened ($t_\text{p}$=3.3\,ns, to supply the same energy) square pulse (black) were measured at the same nominal $V_\text{max} = 0.159\,$V. The amplitude of the square voltage pulse at $t_\text{p}=10\,$ns (gray) is 98\,mV and was chosen to obtain $P_\text{switch}=95\%$. (e) Probability distribution of fitted nucleation times $t_\text{0}$ of 500 single switching events for the respective pulse shapes.}
\end{figure}

For a quantitative overview, we fit a sigmoidal function to the switching probability as a function of $V_\text{max}$ and $E_\text{p}$ and derive the critical switching voltage $V_\text{c}$ and energy $E_\text{c}$ where the fit crosses $P_\text{switch} = 50\%$.  
Figure~\ref{fig2:PP_shapes}(e) shows $E_\text{c}$ as a function of $t_\text{p}$. The square pulse is the least energy efficient for any $t_\text{p}$. For $t_\text{p} =10$~ns, the descending triangular and half cycle sine wave enable magnetization reversal at about 38\% reduced energy at $P_\text{switch}=50\%$ with respect to the square pulse. The ascending triangular pulse and the symmetric triangular pulse provide a better energy efficiency, with an energy reduction of 50\% as compared to the square pulse. Similarly, for $t_\text{p} =1$\,ns, the square pulse is the least efficient, whereas the symmetric triangular pulse performs best, followed by the sine and descending triangular pulse. The ascending triangular pulses have become worse compared to the 10\,ns case. \mh{This behavior showcases the importance of the central time interval of the pulse with respect to the end of the pulse for short pulses, which originates from the interplay of heating and SOT effects (see Sect.~\ref{section_voltage_spikes})}. A weak minimum in $E_\text{c}$ can be observed around 1~ns, as expected due to the transition from the intrinsic to the thermally-activated switching regime \cite{Sala2022}. For sub-nanosecond pulses, the different shapes become blurred due to the finite rise time and bandwidth of the experimental setup. Consequently, the differences in switching energy also tend to vanish. 

\subsection{Time-resolved measurements}
In order to time-resolve the change of $R_\text{MTJ}$, we applied $V_\text{MTJ} = 1.75\,V_\text{SOT}$ which induces both STT and VCMA. The selected polarity of $V_\text{MTJ}$ supports the P-AP switching and corresponds to about 30\% of the required pure STT switching voltage.
The voltage transmitted during the pulse is normalized with respect to the averaged voltage traces in the initial and final state to yield the normalized $\Delta V_\text{sw}$ \cite{Grimaldi2020}. This quantity then reflects the time-dependent changes in the $z$-component of the free layer's magnetization. 

The input voltage and normalized time-resolved switching traces averaged over several hundreds of switching events are shown in Fig.~\ref{fig3:TR_shapes}(a,b) and (c,d), respectively, for square and triangular pulses. The switching traces $\Delta V_\text{sw}$ consist of an initial flat part during which the magnetization of the free layer is at rest and a steadily increasing part during which the magnetization reverses. The duration of these two parts define\mh{s} the average nucleation time $t_\text{0}$ and transition time $\Delta t$, respectively. In agreement with previous studies \cite{Baumgartner2017, Grimaldi2020, Krizakova2021, Sala2023}, we assign $t_\text{0}$ to the time required to nucleate an inverted domain and $\Delta t$ to the time required to expand the domain until full reversal is achieved. 

First, we study the switching behavior obtained by square pulses, as shown in Fig.~\ref{fig3:TR_shapes}(a). \mh{The square pulse with $t_\text{p}=10$~ns and amplitude 98~mV (gray line) achieves 95\% switching probability. We note that this pulse supplies a larger energy than the triangular pulses shown in Fig.~\ref{fig3:TR_shapes}(b). A square pulse with reduced amplitude at fixed $t_\text{p}=10$~ns with the same energy as the triangular pulses does not result in reliable switching.} The $\Delta V_\text{sw}$ corresponding to such a pulse, shown in Fig.~\ref{fig3:TR_shapes}(c), reveals that, on average, the reversal phase starts after 3~ns and spreads out over nearly half of the pulse width. 
The shorter square pulse with $t_\text{p}=3.3$~ns and amplitude 159~mV (black line), on the other hand, shortens the reversal to well below 2~ns. 
This pulse provides the same energy as the 10\,ns-long triangular pulses shown in Fig.~\ref{fig3:TR_shapes}(b) and achieves 99.6\% switching probability, but within a shorter $t_\text{P}=3.3$~ns. 

For the three triangular pulse shapes shown in Fig.~\ref{fig3:TR_shapes}(b), the nominal maximum voltage is $V_\text{max} = 0.159$\,V resulting in over-critical switching ($P_\text{switch} \geq 99.9\%$). In this condition, the nucleation of the inverted domain takes place around the peak of the pulse, as is evident from Fig.~\ref{fig3:TR_shapes}\mh{(d)}. 
The switching trace induced by the descending triangle resembles the trace of the short square pulse and completes effectively in the same time (2\,ns) at lower power. The larger noise at the beginning and/or end of $\Delta V_\text{sw}$ observed for the triangular pulses is related to the lower current passing through the MTJ in the ascending/descending part of the pulse, which is proportional to the detected signal. \mh{Finally, real-time measurements of half cycle sine pulses (not shown) yield switching traces comparable to the symmetric triangle, indicating a similar switching dynamics.}

To resolve the switching speed in single-shot reversal events, the normalized voltage traces of individual pulses are fitted by sigmoidal functions. Then, $t_\text{0}$ is defined as the time at which 10\% of the switching is completed and $\Delta t$ as the difference between the times corresponding to 90\% and 10\% of the switching amplitude. 
The statistical distributions of $t_\text{0}$ for different pulse shapes are presented in Fig.~\ref{fig3:TR_shapes}(e). In the case of the descending triangle, the reversal starts just after the pulse onset with a mean and standard deviation of $0.5 \pm 0.3~$ns. 
The distribution of $t_\text{0}$ is mostly symmetric around the mean. The ascending triangle, on the other hand, initiates the reversal much closer to the end of the pulse with a mean of $7.5\pm0.6~$ns. 
Interestingly, the distributions of $t_\text{0}$ for the ascending and the symmetric triangular pulse are negatively skewed. This indicates that a considerable portion of switching events is initiated before the most probable nucleation time, due the slightly over-critical conditions, thermal activation and minor contributions from VCMA and STT.
The reversal of the square pulse with $t_\text{p} = 3.3$\,ns (equal energy as the triangular pulses) starts at around $0.5\pm0.2$~ns, similar to the descending triangle. 
On the other hand, square pulses of lower amplitude with $t_\text{p} = 10$\,ns present a much broader and delayed $t_\text{0}$ distribution, with a mean value of $4.8\pm1.9$~ns and the successful switching attempts spread over the entire pulse length. This explains why the averaged switching trace in Fig.~\ref{fig3:TR_shapes}\mh{(c)} appears slower compared to $t_\text{p} = 3.3$\,ns. 
\\

The mean values and standard deviations of $\Delta t$ for all pulse shapes (not shown) do not reveal significant variations and fall into the range $1.0\pm0.2\,$ns. Therefore, the domain wall propagation behavior is not largely influenced by the pulse shape. \mh{This observation is consistent with prior work on 3-terminal MTJs, where only weak changes of $\Delta t$ were reported as a function of $V_\text{SOT}$ and $V_\text{MTJ}$ \cite{Grimaldi2020,Krizakova2020,Krizakova2021}. This behavior can be explained if the domain wall propagation velocity is already close to saturation for the range of $V_\text{SOT}$ required to trigger domain nucleation and initiate the reversal process \cite{Martinez2014, Mikuszeit2015,Baumgartner2018}. We also note that resolving differences in $\Delta t$ below 0.5~ns is hindered by the limited bandwidth of our setup.
\\ 
Overall, the statistical distributions of $t_\text{0}$ and $\Delta t$ indicate that, for the same $E_\text{pulse}$, pulse shaping can lead to significant adavantages not only in terms of energy efficiency but also in terms of switching speed. The different efficiencies and $t_\text{0}$ distributions for the above-presented pulse shapes can be explained by the temperature-activated nature of the domain nucleation process and the different timescales of Joule heating and SOT. Whereas the strength of the torque varies simultaneously with the applied voltage, the temperature increase varies as $\Delta T \propto 1-e^{-t/\tau}$, where $\tau \approx 1.5-4.0$~ns is a thermal time constant that depends on device size \cite{Grimaldi2020,Krizakova2021, Chavent2016, Papusoi2008}. Maximum efficiency and speed of the reversal process are achieved when $\Delta T$ and the SOT peak at approximately the same time. Our modelling in Sect.~\ref{5} further supports this interpretation, showing that the symmetric and ascending triangular pulses lead to higher device temperatures coinciding with larger SOT. In the next section, this reasoning is further exemplified by studying the importance of each interval of the pulse in a systematic way using square voltage pulses with superimposed delayed voltage spikes.}

\section{\label{4} Switching efficiency of square pulses with delayed voltage spikes} 
\label{section_voltage_spikes}
\subsection{Post-pulse measurements}
\begin{figure}
\includegraphics[width=85mm]
{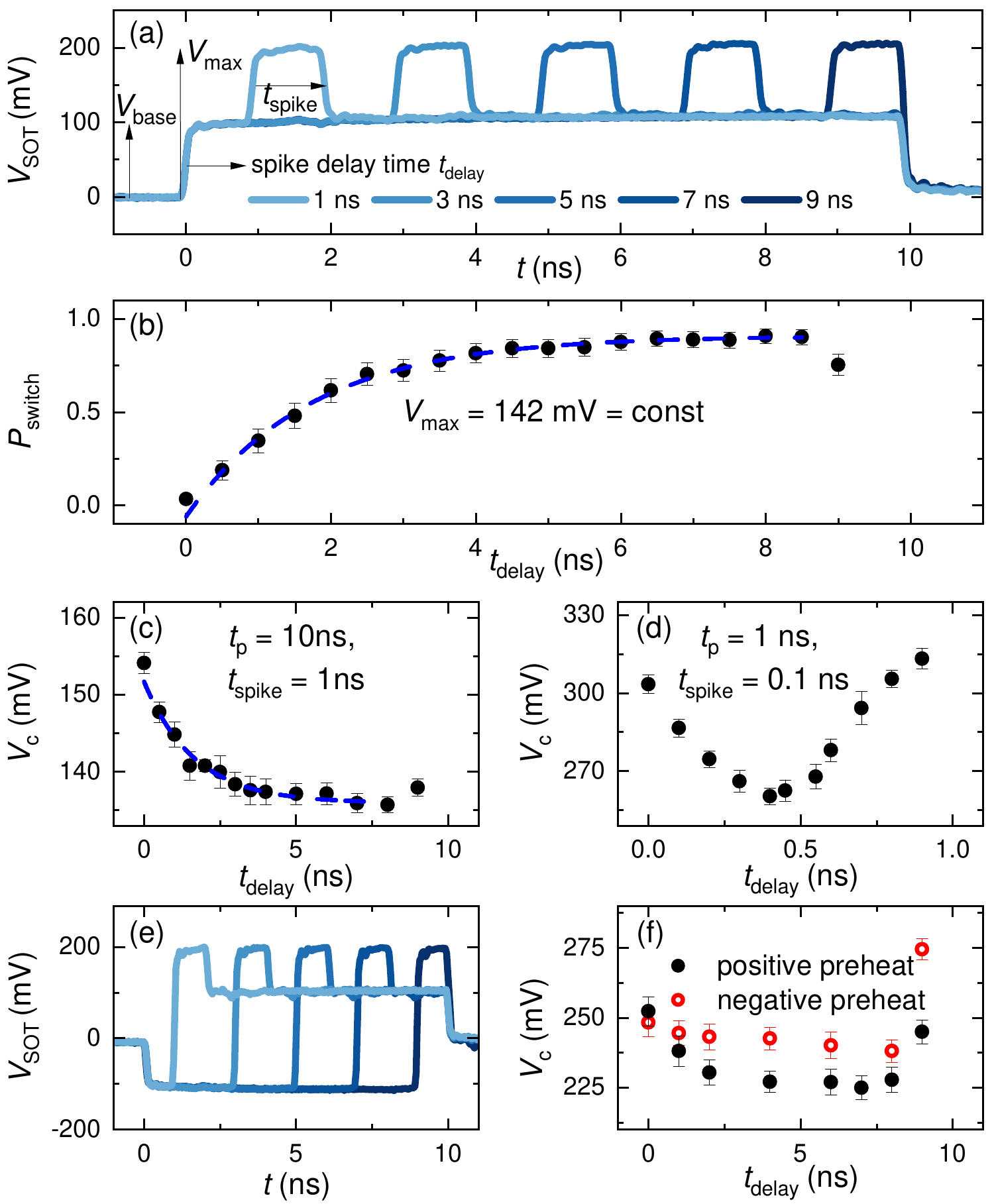}
\caption{\label{fig4:PP_Spikes}  
(a) Pulse shapes consisting of a square baseline ($V_\text{base}$) with an overlapping voltage spike of duration  $t_\text{spike} = 1\,$ns and amplitude $V_\text{max} = 2\,V_\text{base}$ delayed by $t_\text{delay}$. Pulses of total duration $t_\text{p} = 10$\,ns and varying $t_\text{delay}$ are shown. (b) Switching probability as a function of spike delay at a fixed applied voltage fit by an exponential function shown as dashed line. (c) Critical applied maximal voltage $V_\text{c}$ vs $t_\text{delay}$ as determined from the sigmoid curve fits to the voltage sweeps (not shown) and the exponential fit (dashed line) (d) $V_\text{c}$ for the same pulse shape with $t_\text{p} = 1$\,ns, $t_\text{spike} = 0.1$\,ns, and $V_\text{spike} = 2\,V_\text{base}$  for different $t_\text{delay}$. (e) Pulse shapes with a negative preheating phase before the spike. Here, $t_\text{delay}$ marks the time at which the polarity of the voltage switches. (f) Comparison of the $t_\text{delay}$-dependence of $V_\text{c}$ for a preheating phase with supportive (black) and opposing (red) SOT in a different device with modified critical conditions. All data were recorded for the P-AP switch at $V_\text{MTJ} = 0$ and $B_x=-45$\,mT.}
\end{figure}

To understand the importance of different segments of the pulses with total duration of 10~ns or 1~ns, we study here the switching characteristics of two superposed square pulses with a variable delay. We used pulses with a square base amplitude $V_\text{base}$ and length $t_\text{p}$ featuring a square spike of double the base amplitude ($V_\text{max } = 2 \, V_\text{base}$) and duration $t_\text{spike} = 0.1 t_\text{p}$. The delay $t_\text{delay}$ of the spike with respect to the base pulse was varied between 0 and $0.9~t_\text{p}$ [Fig.~\ref{fig4:PP_Spikes}(a)]. In this manner, the total injected angular momentum and pulse energy is constant for different $t_\text{delay}$. \\
At $t_\text{p} = 10~$ns and fixed voltage amplitude, the pure SOT switching probability measured over 500 events increases for longer $t_\text{delay}$ [Fig.~\ref{fig4:PP_Spikes}(b)]. This change is most pronounced during the first few nanoseconds and saturates for longer $t_\text{delay}$. An exponential fit of $P_\text{switch}$ yields a characteristic timescale of $2.0\pm0.3$\,ns resulting in 95\% switching probability after 6\,ns. \mh{This timescale is similar to the self-heating relaxation time reported in MTJ devices \cite{Papusoi2008, Chavent2016, Grimaldi2020}.} However, if the spike ends simultaneously with the base pulse, a decrease in switching probability is noticeable. As $\Delta t$ is about as long as the remaining pulse time after the spike in this case, the magnetization reversal cannot complete reliably. Also $V_\text{c}$ is influenced by $t_\text{delay}$ [Fig.~\ref{fig4:PP_Spikes}(c)], decreasing exponentially with $t_\text{delay}$ until $t_\text{p} - t_\text{spike} = 9\,$ns, for which a slight increase in $V_\text{c}$ is observed, consistently with the reduction of $P_\text{switch}$ reported above. This trend is in line with the increased energy efficiencies of the symmetric and ascending triangular pulses described in Fig.~\ref{fig2:PP_shapes}(c). \\
In contrast, at $t_\text{p} = 1\,$ns $V_\text{c}$ in Fig.~\ref{fig4:PP_Spikes}(d) has a minimum approximately at the center of the pulse. In accordance with Fig.~\ref{fig2:PP_shapes}(d), these results demonstrate that the free layer is easiest to switch when the highest driving voltages appear neither in the beginning nor in the end of the pulse. \mh{This behavior is consistent with the requirement of a preheating phase leading to thermally-activated domain nucleation and a subsequent domain expansion phase driven by SOT.}
\\ To distinguish between the action of torque and self-heating, we further investigate a preheating phase of duration $t_\text{delay}$ with inverted voltage $-V_\text{base}$ and inverted torques [Fig.~\ref{fig4:PP_Spikes}(e)]. The resulting $V_\text{c}$ \textit{vs}. $t_\text{delay}$ is presented for a preheating phase with either supportive and opposing SOT in Fig.~\ref{fig4:PP_Spikes}(f). Whereas the preheating phase with opposing torque qualitatively resembles the preceding results, the decrease of $V_\text{c}$ is not as pronounced as for the supportive torque. Thus, the switching success of the voltage spike is not solely given by the self-heating effect of the preheating phase. Instead, the difference of the two data sets in Fig.~\ref{fig4:PP_Spikes}(f) implies that the SOT also plays an important role in triggering the reversal.

\subsection{Time-resolved measurements}
\begin{figure}
\includegraphics[width=85mm]{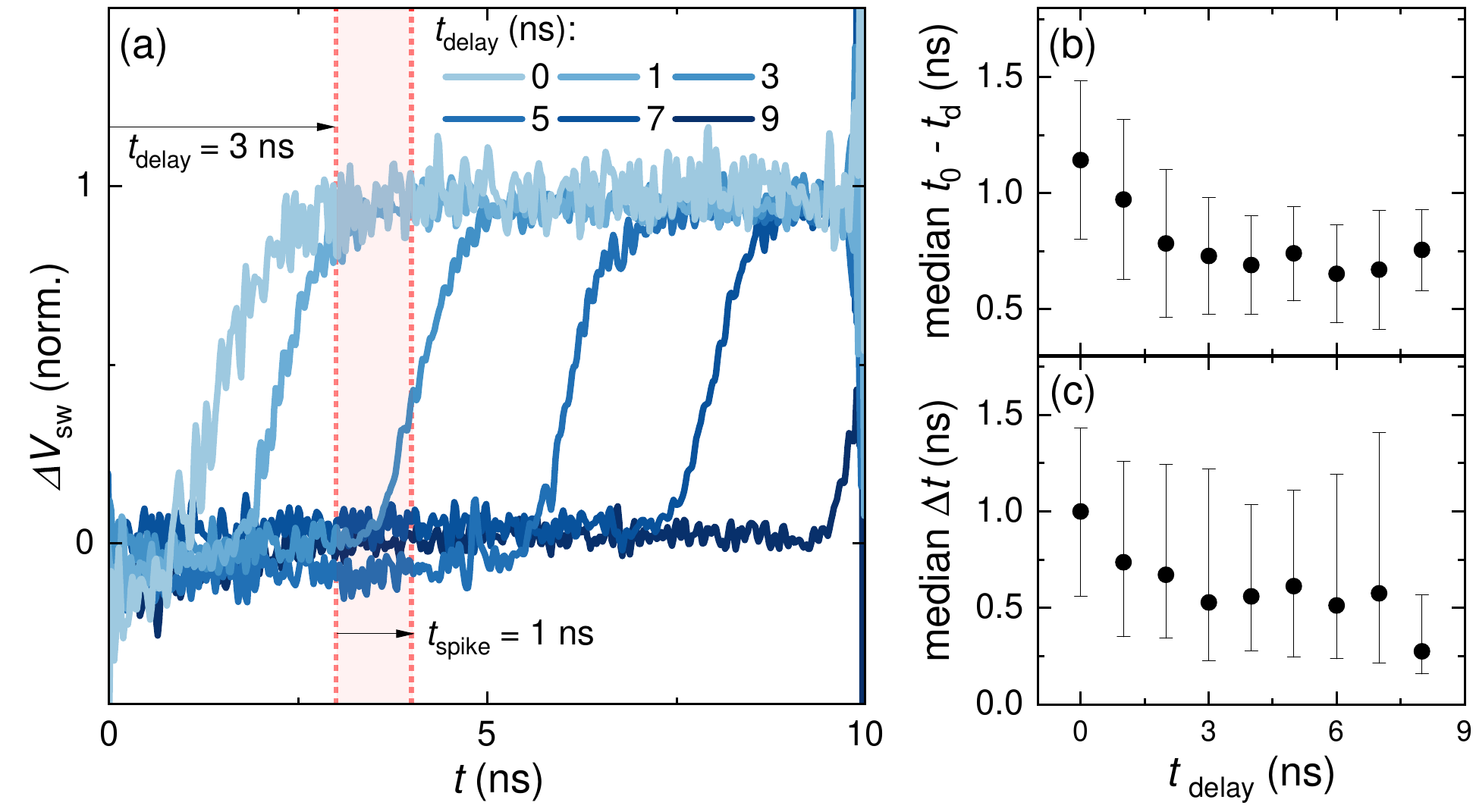}
\caption{\label{fig5:TR_Spikes} (a) Time-resolved switching traces for the pulse shapes presented in Fig.~\ref{fig4:PP_Spikes}(a) averaged over $\approx 500$ events. The shaded area indicates the duration of the spike with $t_\text{delay}=3\,$ns. (b) Dependence of the median nucleation time $t_\text{0}$ offset by the corresponding $t_\text{delay}$ and (c) $\Delta t$ on $t_\text{delay}$ , as determined by sigmoid fits to the individual switching events. The error bars span lower and upper quartiles of the distributions. All data were recorded at $V_\text{MTJ} = 2\,V_\text{SOT}$, $t_\text{p}=10\,$ns, and $t_\text{spike}=1\,$ns.} \end{figure}

To investigate whether the discontinuity caused by the voltage spike makes the two-phase reversal process less uniform, time-resolved measurements were performed: 
Figure \ref{fig5:TR_Spikes}(a) presents averaged switching traces for the pulse shapes shown in Fig.~\ref{fig4:PP_Spikes}(a) at constant $V_\text{max} = V_\text{spike} =130\,$mV and $V_\text{MTJ} = 2 \, V_\text{SOT}$. Due to the trend illustrated in Fig.~\ref{fig4:PP_Spikes}(b), these conditions are undercritical for $t_\text{delay}$ = 0~ns yielding a switching probability of 15\%.
For longer $t_\text{delay}$, reliable switching is achieved. For higher driving voltages, the magnetization dynamics would be accelerated and the differences between traces more subtle.

The averaged switching traces reveal close similarities with the ones presented in Fig.~\ref{fig3:TR_shapes}(d): The part of the pulse featuring the largest voltage amplitudes, i.e.~the spikes here, trigger the reversal process in critical conditions. The reversal initiates on average within the spike. Therefore, the switching traces appear to be offset in time approximately by the respective differences in $t_\text{delay}$, but are otherwise similar to each other.
A mild trend is revealed by analyzing how the difference $t_\text{0} - t_\text{delay}$ depends on $t_\text{delay}$. This quantity measures how soon the spike succeeds in initiating the reversal. 
The medians of $t_\text{0} - t_\text{delay}$ and $\Delta t$ determined by fits to the successful individual switching events that complete within $t_\text{P}$ are presented for several $t_\text{delay}$ in Fig.~\ref{fig5:TR_Spikes}(b) and (c), respectively. For longer $t_\text{delay}$, we find shorter intervals between spike onset and nucleation time. This decrease reflects the same trend as observed for $V_\text{c}$ in Fig.~\ref{fig4:PP_Spikes}(c) and may be understood by considering that $t_\text{0}$ is a measure of the energy barrier that needs to be overcome in the reversal process \cite{Sala2023}. The energy barrier and thus $t_\text{0}$ are diminished thanks to the higher temperature reached for longer $t_\text{delay}$. 
A similar, even more subtle trend is obtained for the medians of $\Delta t$ [cf.~Fig.~\ref{fig5:TR_Spikes}(c)]. Therefore, higher temperatures appear to enhance the reversal speed, consistently with models of domain wall propagation in the presence of a pinning potential \cite{Sala2023}, leading to shorter $\Delta t$.
Overall, also the time-resolved switching data evidence the influence of temperature, supporting the scenario in which pulse spikes occurring after a preheating phase of a few nanoseconds facilitate the switching.

\section{\label{5} Micromagnetic model including self-heating} 
To incorporate the temperature dynamics induced by different pulse shapes into the micromagnetic model, we assume that the saturation magnetization $M_\text{S}$ scales with temperature as $M_\text{S} = M_\text{S,0} \left( 1-T/{T_\text{c}}\right)^{\beta}$, where $M_\text{S,0}=1.83$~MA/m, $T_\text{c}=750$~K is the Curie temperature, and $\beta=1$ the critical exponent. 
The magnetic anisotropy $K_\text{u}(T)$ (845~kJ/m$^3$ at 300~K) and exchange stiffness $A_\text{ex}(T)$ (15 pJ/m at 300~K) 
scale like $K_0\left(\frac{M_{\mathrm{S}}(T)}{M_{\mathrm{S} 0}}\right)^p$ and $A_{\mathrm{ex} 0}\left(\frac{M_{\mathrm{S}}(T)}{M_{\mathrm{S} 0}}\right)^q$ with $p$ = 2.5 and $q$ = 1.7, respectively.  
More details are found in the literature \cite{Lee2017a, Grimaldi2020}. The final temperature induced by self-heating is proportional to the square of the current density \cite{Sousa2004}, assuming a negligible change in resistance. The system requires a finite time to equilibrate the temperature while heat is dissipated via the adjoining layers and the material surrounding the MTJ pillar. 
We describe the continuous release of heat into the environment by an exponential decay with an effective relaxation time $\tau = 2.5\,$ns \cite{Grimaldi2020, Kim2008} convoluted with self-heating:
\begin{equation}
\label{equ_jouleheating}
    c_\text{eff}\, \Delta T(t)  =  \int_0^t R_\text{SOT} A_\text{SOT}^2 j_\text{SOT}^2(t') e^{-\frac{t-t'}{\tau}} dt',
\end{equation}
where $ c_\text{eff}$ and $A_\text{SOT}$ are the effective heat capacity of the MTJ and the cross-section of the SOT track, respectively, and $j_\text{SOT}(t)$ is the injected current density.  

\begin{figure}
\includegraphics[width=85mm]{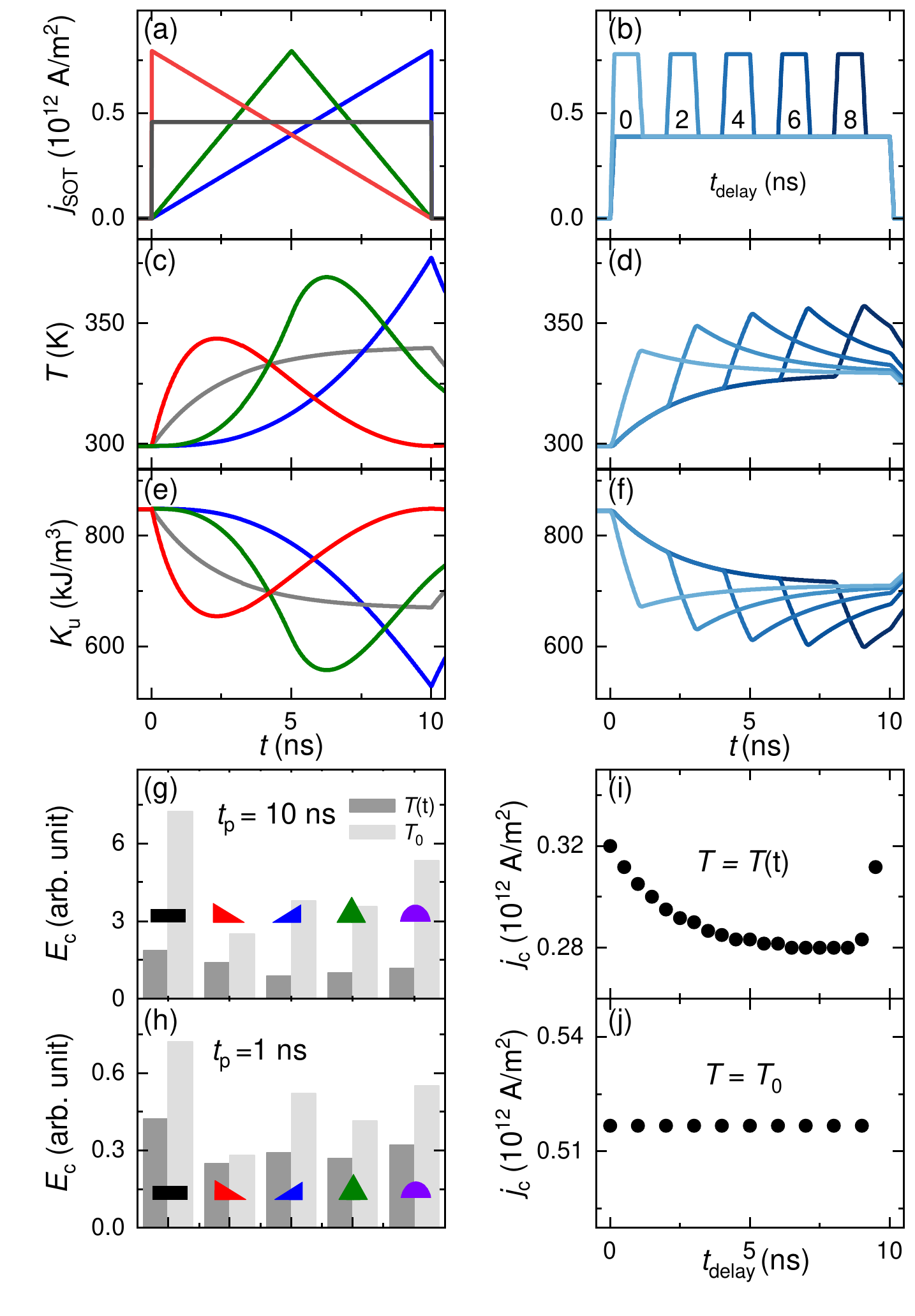}
\caption{\label{fig7_sims} \mh{(a) Single pulses and (b) spiked pulses with varying $t_\text{delay}$ together with their corresponding transient evolution of (c, d) temperature and (e, f) uniaxial anisotropy. Critical switching energy for the different pulse shapes at (g) $t_\text{p} = $10\,ns and (h) 1\,ns as determined from the lowest current density that achieves switching $j_\text{c}$. The simulations are performed for a time-dependent temperature given by Eq.~\eqref{equ_jouleheating} (dark gray) and constant temperature $T_\text{0}= 300\,$K (light gray). Critical current density $j_\text{c}$ for the square pulse shapes with amplitude spikes as a function of $t_\text{delay}$ (i) with and (j) without temperature variation.}}
\end{figure}

\mh{Using this approach, we simulate pure SOT switching induced by the pulse shapes shown in Fig.~\ref{fig7_sims}(a) and (b). The resulting transient dependencies of temperature and $K_\text{u}$ are given in Fig.~\ref{fig7_sims}(c-f).}
By varying $j_\text{SOT}$, we extract the lowest amplitude $j_\text{c}$ at which successful switching occurs. 
\mh{For different pulse shapes, $j_\text{c}$ can be converted to the corresponding energy $E_\text{c}$ using Eq.~\eqref{equ_E-V} for $t_\text{p} = $10~ns and 1~ns, as shown in Fig.~\ref{fig7_sims}(g) and (h), respectively. The simulation results for $E_\text{c}$ and the different pulse shapes are plotted considering a constant and a variable temperature scenario.} Generally, we find good agreement with the energy efficiency of the pulse shapes investigated experimentally [Fig.~\ref{fig2:PP_shapes}(e)]. In the thermally-activated regime ($t_\text{p} = $10~ns), the reduction in $E_\text{c}$ is more pronounced than for 1~ns. The descending triangle provides its largest SOT instantly after the onset of the pulse. Therefore, it benefits the least from the temperature increase that follows only at later times.

\mh{The simulated $j_\text{c}$ for the pulses shown in Fig.~\ref{fig7_sims}(b) are plotted as a function of $t_\text{delay}$ in Fig.~\ref{fig7_sims}(i). These values reflect the experimental observations with the exponentially saturating decrease of $j_\text{c}$ due to preheating and the mild increase due to the finite reversal completion time.}
Without the transient variation of temperature and anisotropy, the variation of $j_\text{c}$ with $t_\text{delay}$ disappears, as shown in  Fig.~\ref{fig7_sims}(j). 
\mh{Overall, the simulations corroborate our interpretation of the experimental trends and confirm that the duration of the preheating phase and of the domain wall propagation phase determine the optimal pulse shape for a given maximum current amplitude.}

\section{Conclusions}
We investigated the SOT switching induced by shaped voltage pulses in three-terminal MTJs using post-pulse and time-resolved measurements as well as micromagnetic simulations.
We found that modulating the voltage amplitude over the pulse duration can induce substantial gains in energy efficiency without necessarily degrading the switching speed. In the thermally-activated regime at $t_\text{p}$ = 10\,ns, the ascending and symmetric triangular pulse are the most efficient, indicating large voltage amplitudes between the center and the end of the pulse favor the reversal process. This finding is rationalized by a preheating effect: the initial part of the pulse raises the device temperature and lowers the domain nucleation barrier, followed by a voltage increase that induces a strong SOT and achieves the nucleation. Upon approaching the intrinsic regime at $t_\text{p}$ = 1\,ns, the symmetric triangular pulse is the most efficient, while the descending triangular pulse is more efficient than the ascending triangle. Hence, it is more favorable to provide large voltages around the center of the pulse. This trend can be explained by considering that, in the intrinsic limit, the strongest voltage should be provided at the beginning of the pulse in order to trigger domain nucleation, followed by a smaller voltage to drive the domain expansion. The observations for both switching regimes are confirmed by systematically shifting a pulse spike on top of a square base pulse.

Time-resolved measurements demonstrate that the largest voltage amplitude within a given pulse trigger domain nucleation and initiate the magnetization reversal at critical switching conditions. The statistical distributions of the nucleation time $t_\text{0}$ are strongly affected by the pulse shape. Near the critical switching voltage, the total switching time is dominated by $t_\text{0}$. On the other hand, expanding the nucleated inverted domain through the entire free layer requires a finite time $\Delta t$ (here $\approx 0.8 \pm 0.4$~ns), which is only mildly impacted by changing the pulse shape and temperature of the device. Consequently, the optimal efficiency and switching speed are achieved when the largest voltage amplitude is applied near the middle of the pulse and earlier than $\Delta t$ before the pulse terminates.

Micromagnetic simulations implementing time-dependent SOT proportional to the pulse shape $V(t)$ and a simple dynamic temperature relaxation model based on self-heating reproduce the experimental observations. The simulations performed for different pulse shapes and durations reveal the characteristic timescale on which the device heats up and interplay between SOT and thermally-activated dynamics, which is key to optimize the pulse shape and duration to achieve maximum switching efficiency on the nanosecond timescale.

In practical terms, the switching speed and energy are both optimized by a symmetric triangular pulse saving about 39\% (50\%) with respect to a square pulse lasting 1~ns (10~ns). 
Additional measurements show that a sine-shaped pulse, as commonly used in radiofrequency electronics, has qualitatively similar effects to the symmetric triangle and saves about $25\%$ in switching energy with respect to a square pulse.
Our work shows how shaped pulses can contribute to lowering the critical switching energy of MTJs in memory devices. Future investigations may take advantage of shaped pulses to implement nonbinary switching algorithms, as required, e.g., in probabilistic and neuromorphic computing architectures \cite{Wu2022, Sengupta2016, Zheng2020}.

\begin{acknowledgments}
This project has received funding from the European Union’s Horizon 2020 research and innovation programme under the Marie Skłodowska-Curie grant agreement N. 955671.
This work was further supported by the Swiss National Science Foundation (Grant N. 200020\textunderscore200465) and IMEC's IIAP industrial partner affiliation program on SOT-MRAM.
\end{acknowledgments}

\bibliography{main}

\end{document}